\documentclass[aps,prl,floatfix,twocolumn]{revtex4-1}
\usepackage{graphicx}
\usepackage{epstopdf}
\usepackage{dcolumn}
\usepackage{bm}
\usepackage{setspace}
\usepackage{psfrag}
\usepackage{float} 
\usepackage{color}
\newcommand{\eeq}{\end{eqnarray}}

\begin{document}
\title{Role of Excited States In High-order Harmonic Generation}

\author{S. Beaulieu$^{1,2}$,
S. Camp$^{3}$,
D. Descamps$^{1}$,
A. Comby$^{1}$,
V. Wanie$^{1,2}$,
S. Petit$^{1}$,
F. L\'egar\'e $^{2}$,
K. J. Schafer$^{3}$,
M. B. Gaarde$^{3}$,
F. Catoire$^{1}$
Y. Mairesse$^{1}$} 
\bigskip

\affiliation{
$^1$Universit\'e de Bordeaux - CNRS - CEA, CELIA, UMR5107, F33405 Talence, France\\
$^2$ Institut National de la Recherche Scientifique, Centre EMT, J3X1S2, Varennes, Quebec, Canada\\
$^3$ Department of Physics and Astronomy, Louisiana State University, Baton Rouge, Louisiana 70803-4001, USA\\
}

\date{\today}

\begin{abstract}
We investigate the role of excited states in High-order Harmonic Generation by studying the spectral, spatial and temporal characteristics of the radiation produced near the ionization threshold of argon by few-cycle laser pulses. We show that the population of excited states can lead either to direct XUV emission through Free Induction Decay or to the generation of high-order harmonics through ionization from these states and recombination to the ground state. By using the attosecond lighthouse technique, we demonstrate that the high-harmonic emission from excited states is temporally delayed by a few femtoseconds compared to the usual harmonics, leading to a strong nonadiabatic spectral redshift. 

\end{abstract}
\maketitle

High-order Harmonic Generation (HHG) results from the interaction of a strong laser with atoms or molecules, and can be understood as a three-step mechanism \cite{kuchiev87,corkum93,schafer93}. First, an electron wave packet (EWP) is created by strong-field ionization from the ground state. The EWP is accelerated by the laser. Last, it can radiatively recombine with the ion, leading to the emission of extreme ultraviolet (XUV) photons. This model successfully describes the main characteristics of HHG well above the ionization threshold, where the influence of the ionic potential on the EWP dynamics can often be neglected \cite{lewenstein94}. However, below and near the ionization threshold, or in the presence of resonances, the HHG mechanism becomes more complex.  

A particularly important question is the influence of resonances in HHG. First, resonances in the continuum can induce enhancements of the harmonic emission \cite{ganeev06,strelkov10,rothhardt14},  spectral phase jumps \cite{haessler13,ferre15} and lead to strong polarization-state variations when driven with elliptical light \cite{ferrea.15}.  Below the ionization threshold, bound-bound resonances increase the harmonic emission. Chini \textit{et al.} observed narrow-band enhancement by atomic resonances (Rydberg states) \cite{chini14}. They also demonstrated that the emission showed the same ellipticity dependence as the above-threshold harmonics, and suggested that polarization gating techniques \cite{sola06-1} could thus be employed for temporal confinement. On the other hand, a recent theoretical study demonstrated that the narrow-band emission emerged for long-lived dipoles which coherently emit radiation for times much longer than the pulse duration \cite{camp15}. This process can be seen as XUV Free Induction Decay (xFID) \cite{bloch46,brewer72,beck14,bengtsson15}. The temporal confinement of the emission from Rydberg states thus needs to be clarified. It is of particular importance since it could be useful to produce quasi-circularly polarized ultrashort XUV pulses \cite{ferrea.15}. Last, theoretical works have shown that bound-bound resonances can influence the ionization step in HHG \cite{taieb03,ngoko_djiokap15}. For example, Bian and Bandrauk \cite{bian10,bian13} predicted that the resonant population of excited states during the laser pulse could open a new channel for HHG, \textit{i.e.} ionization from excited states and recombination to the ground state. However, to the best of our knowledge, this effect has never been observed  experimentally.

In this letter, we investigate the role of resonances and excited states in HHG from argon atoms using ultrashort laser pulses. We identify xFID emission associated with the excitation of the $[Ne]3s^23p^6 \rightarrow [Ne]3s^23p^5ns$ and $[Ne]3s^23p^5nd$  Rydberg series. We observe new spectral features which emerge from the ionization of electronically excited states and recombination to the ground state (e-HHG). We investigate the temporal properties of the different XUV emission mechanisms using the attosecond lighthouse technique \cite{vincenti12}. We find that while the below- and above-threshold non-resonant harmonics show clear sub-cycle confinement synchronized with the driving infrared laser field, the xFID shows no sign of attosecond structure. In addition, the e-HHG emission occurs only on the falling edge of the laser pulse.  

The experiments were performed with the 1 kHz AURORE laser system at CELIA which delivers 8 mJ, 28 fs, 800 nm pulses.  We focused up to 4 mJ into a 1.5 m long stretched hollow core fiber (500 $\mu m$ diameter, few-cycle inc.) filled with a pressure gradient of argon (0-400 mbar) to broaden the sprectrum ($\sim$ 650-950nm tail-to-tail) through self-phase modulation. Six pairs of chirped mirrors (-50 $fs^2$ per bounce, Ultrafast Innovations) were used to compensate the group delay dispersion. A single-shot, in-line ultra-broadband second-order autocorrelator (FemtoEasy) was used to measure the pulse duration. The pulse energy was varied by rotating a super-achromatic half-wave plate (Fichou) in front of a broadband polarizer (FemtoLaser). The pulses were sent under vacuum and focused by a f=37.5 cm spherical mirror into a 250 $\mu m$ thick effusive Ar gas jet. The XUV radiation was analyzed by a flat-field XUV spectrometer, consisting of  a 1200 grooves/mm cylindrical grating (Shimadzu) and a set of dual microchannel plates coupled to a fast P46 phosphor screen (Hamamatsu) enabling single shot measurements. A 12-bit cooled CCD camera (PCO) recorded the spatially-resolved harmonic spectra. 

\begin{figure}
\begin{center}
\includegraphics[width=7 cm,keepaspectratio=true]{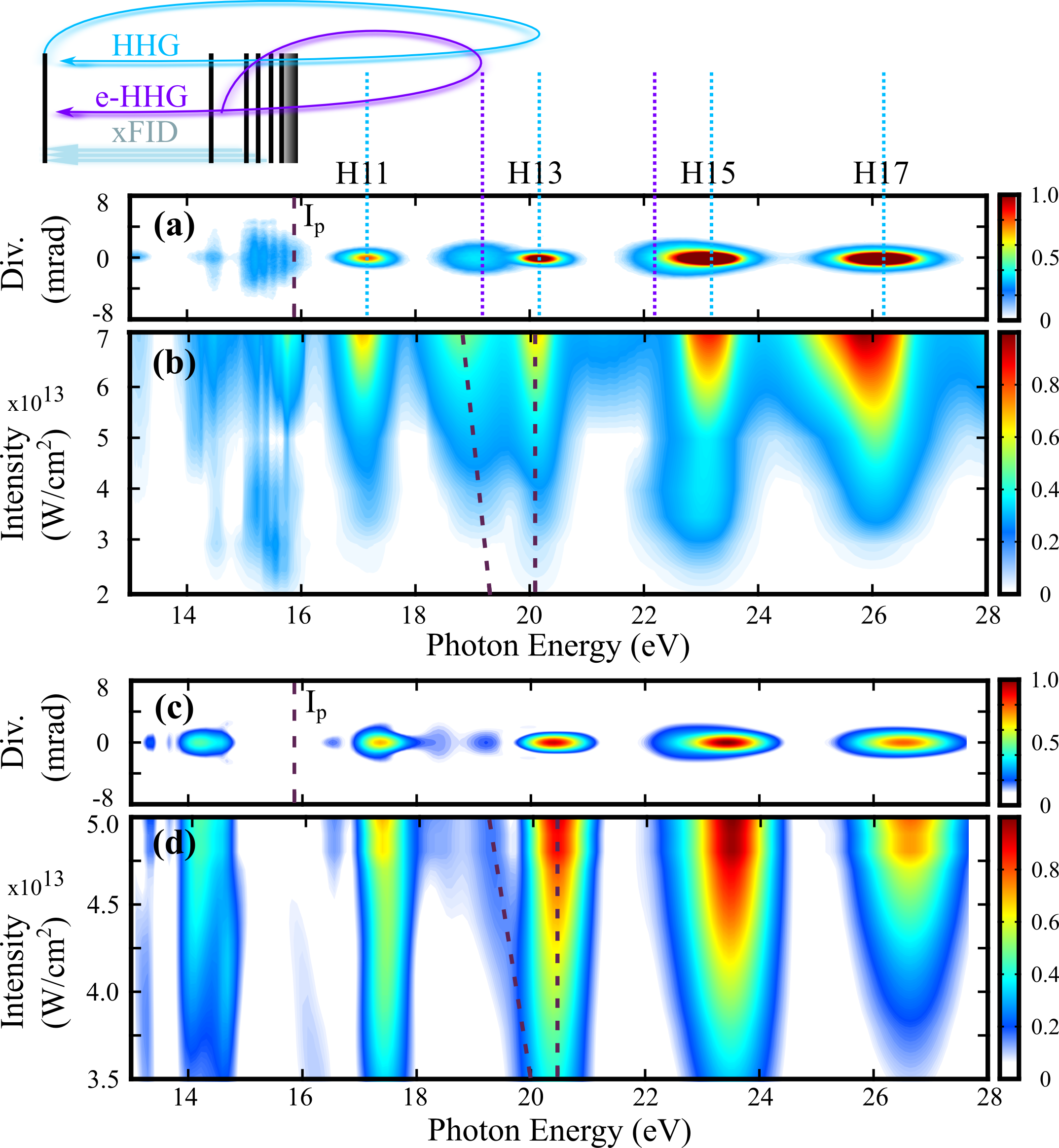}
\caption{(a) Spatially resolved HHG spectrum using 7 fs laser pulses ($I_0$ = $5.2$ x $10^{13}$ $W/cm^2$). (b) Intensity dependence of HHG spectra using 7 fs pulses. (c) TDSE calculation of spatially resolved HHG spectrum using 7 fs laser pulse ($I_0$ = $5.0$ x $10^{13}$ $W/cm^2$). (d) TDSE calculation of the intensity dependence of HHG spectra. The dashed line serves as guides to the eyes. }
\label{fig1}
\end{center}
\end{figure}

Figure \ref{fig1}(a) shows the spatially resolved HHG spectrum of argon driven with 7 fs pulses. In addition to the usual frequency comb separated by twice the laser frequency, new structures appear in the harmonic spectrum : a series of narrow spectral lines between 14.2 and 15.6 eV, a broad peak around 19 eV near H13 and a distinct shoulder on the low-energy side of harmonic 15 around 23 eV. Varying the laser intensity from $2.0$ to $7.0 \times 10^{13}$ W/cm$^2$) (Fig. \ref{fig1}(b)) shows that the narrow-bandwidth structures appear at very low intensity and do not exhibit a shift in energy with intensity. The peak around 19 eV is only visible when the laser intensity is above $3.5 \times 10^{13}$ W/cm$^2$, and shows a linear redshift as the intensity increases, with a slope of $\sim -1.2 \times 10^{-11}$ meV/W $\cdot$ cm$^2$. 

What is the origin of these two components? The narrow structures below the ionization threshold are clearly associated to bound-bound emission from Rydberg states, as recently predicted \cite{camp15}. Because the excitation mechanism is coherent, all atoms in the medium decay in phase to the ground state by spontaneous emission, resulting in coherent emission referred to here as xFID. xFID has recently been identified in the case of excitation of Rydberg and Fano resonances by an XUV beam \cite{bengtsson15}. The emission process is similar here, the difference being that multiple IR photons are absorbed to populate the excited states. The coherence of the xFID process explains the collimated nature of the emission. Since the lifetime of the Rydberg states is much longer than the IR pulse duration, the xFID mostly takes place after the laser pulse is over, at photon energies corresponding to the field-free resonance energies \cite{camp15}. 

\begin{figure}
\begin{center}
\includegraphics[width=8cm,keepaspectratio=true]{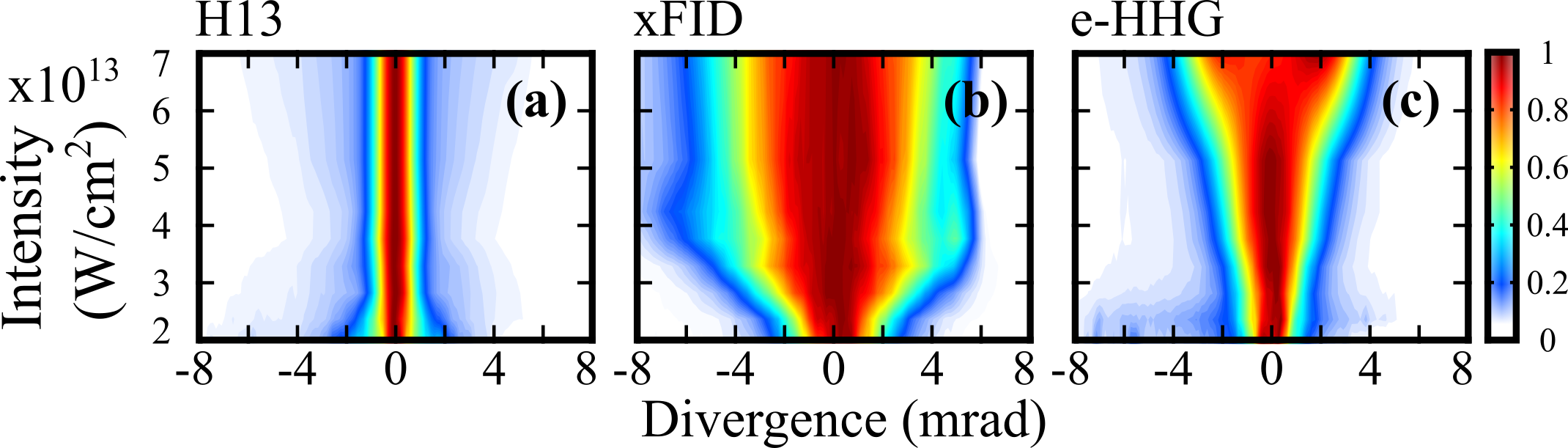}
\caption{Intensity dependence of the divergence for (a) H13 ($\sim$20 eV), (b) xFID ($\sim$15 eV) and (c) e-HHG ($\sim$19 eV). Note that the observed saturation of the xFID divergence in (b) is due to a limited numerical aperture in the experiment.}
\label{fig2}
\end{center}
\end{figure}

The origin of the feature located $\sim$3.2 eV above the $I_p$ ($\sim$19 eV) is not as obvious as that of the xFID features. Its spectral shift with intensity indicates that it occurs when the laser field is on. Interestingly, it shows a broader spatial profile than the neighboring harmonics. We investigate this spatial profile disparity as a function of laser intensity in Fig. \ref{fig2}. In conventional HHG, the spatial profile of harmonic $q$ is dictated by the curvature of the atomic dipole phase ($\phi _q ^ {l,s} (r,t)$) in the generating medium. This phase depends linearly on the laser intensity profile, $I_0 (r,t)$: $\phi _q ^ {l,s} (r,t) = - \alpha _ q ^{l,s} \cdot I_0 (r,t)$. The coefficient $\alpha _ q ^{l,s}$ is determined by the electron trajectory in the continuum, and its value increases with the trajectory excursion time. Each harmonic $q$ can be emitted by two different electron trajectories, labeled short ($s$) and long ($l$), with respectively a small and large $\alpha$, a small and large spatial divergence, and a slow and fast increase of the spatial divergence with laser intensity. We used the spatial profile to retrieve the value of $\alpha$ of each spectral component and gain some insight about the associated electron dynamics in the continuum. The procedure is detailed in the SI. For e-HHG, this analysis provides  $\alpha = (14 \pm 1) \cdot 10 ^ {-14} cm^2/W $, which corresponds to the emission of long trajectories near the cutoff \cite{gaarde02,varju05}. Thus, these results suggest that the e-HHG could be emitted by a channel with a much lower cutoff than the ground state emission from argon. We performed the same analysis for the xFID (\ref{fig2} (c)) emission and found a much larger value,  \textit{i.e.} $\alpha = (33 \pm 5) \cdot 10 ^ {-14} cm^2/W$. This physical quantity gives us information about the effect of the driving laser intensity on the initial phase of the xFID. 

To gain more insight into the role of excited states in the emission, we performed Time-Dependent Schr\"odinger Equation (TDSE) calculations. First, the single atom response was calculated solving the 1D-TDSE in the velocity gauge, using a soft core potential with the asymptotic Coulomb tail and argon ionization potential. The calculation was performed using a 7 fs FWHM gaussian pulse, and repeated for different peak intensities. Second, the macroscopic signal was calculated by defining a gaussian spatial distribution of the laser intensity. The spatially resolved far-field harmonic spectrum was obtained by calculating the Hankel transform of the near-field dipole distribution. This calculation neglects longitudinal phase matching effects, which is justified by the thin nature of the gas jet used in the experiment \cite{catoire16}. Figure 1(c) shows the calculated spatially resolved HHG spectrum at $5 \cdot 10^{13} W/cm^2$ and Fig. 1(d) represents the HHG spectrum as a function of the laser intensity. Remarkably, the structure around 19 eV and the broadening of the red wing of H15 show up in the simulated spectrum. As in the experimental results, we see that the signal around 19 eV experiences a spectral redshift as the laser intensity increases. The slope of the energy shift is $\sim -2.5 \times 10^{-11}$ meV/W $\cdot$ cm$^2$. We performed the same simulation using a screened Coulomb potential which does not support any excited state, and found that these spectral features vanished. This observation confirms the key role of excited states in the mechanism. A further analysis shows that the main contribution in the microscopic dipole stems from the recombination to the ground state. This means that excited states are involved in the ionization rather than recombination step.

\begin{figure}
\begin{center}
\includegraphics[width=7 cm,keepaspectratio=true]{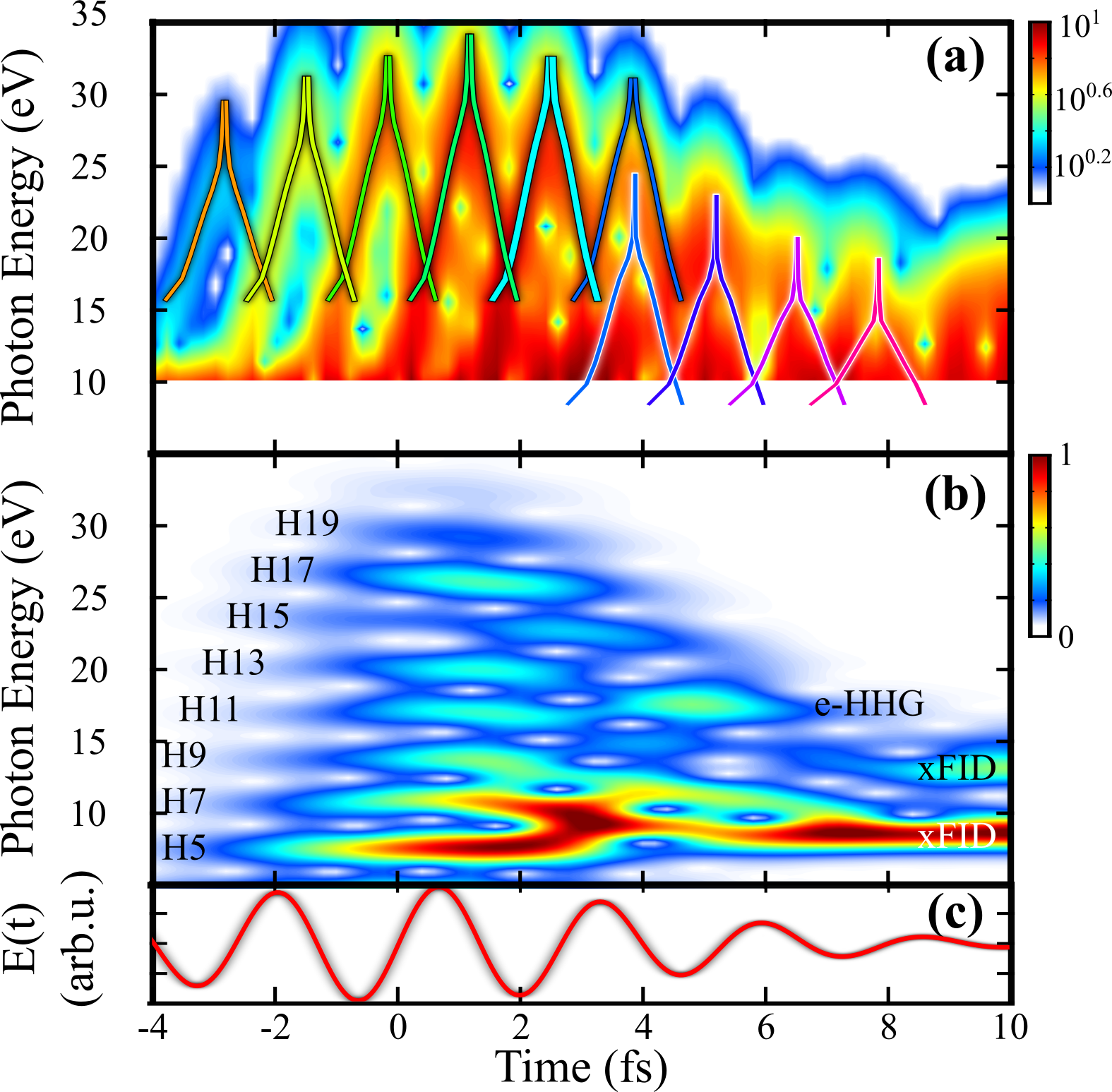}
\caption{Gabor analysis of the TDSE dipole, at $I_0$ = $5.0 \times 10^{13}$ $W/cm^2$ (a) Gabor analysis using a 0.2 optical cycle FWHM Gaussian window function. The colored branches represent the emission times from short- and long- quantum trajectories calculated with SFA. The upper branches (from -3 fs to 4 fs) correspond to both ionization and recombination from/to the ground state. The lower branches (4 fs to 10 fs) involve ionization from the excited state and recombination to the ground state. (b) Gabor analysis using a 0.75 optical cycle FWHM Gaussian window function (c) Laser electric field used in the calculations.}
\label{fig3}
\end{center}
\end{figure}

The dynamics of the HHG process can be revealed by performing a time-frequency Gabor analysis of the dipole obtained with the TDSE simulation. First, we investigate the attosecond structure of the emission by using a short Gaussian window function with 530 as FWHM (Fig. \ref{fig3} (a)). During the rising edge of the laser pulse, the spectrogram nicely reveals branches associated to the emission of attosecond pulses from short and long electron trajectories, with respectively a positive and negative slope (attochirp \cite{varju05}). These branches can be compared to the emission times obtained by solving the saddle point equations in the Lewenstein model of HHG (Strong Field Approximation \cite{lewenstein94}), shown as lines superimposed on the spectrogram. Up to $\sim$ 4 fs, the TDSE and SFA results are in remarkable agreement. However, after $\sim$ 4 fs, lower energy components show up, and dominate the emission. To investigate the role of the excited states in the ionization step, we calculated the SFA 
quantum trajectories for the scenario where the ionization takes place from the excited state at 8.9 eV followed by the recombination on the ground state (e-HHG, short for excited states HHG). This process produces an additional comb of high order harmonic shifted in energy by 8.9 eV in agreement with TDSE calculations. Since, in this situation, the ionization potential is reduced to 6.9 eV, the cutoff of this second harmonic spectra appears around 20 eV.  Indeed, the component emitted around 19 eV at $\sim$ 4 fs can be assigned to the cutoff of the e-HHG process, which is consistent with the experimental observation of a large $\alpha$.

In order to resolve the energy of the harmonics, we increased the duration of the Gabor window to 1.6 fs. Figure \ref{fig3} (b) reveals that the emission around 19 eV in the falling edge of the pulse is shifted in frequency with respect to neighboring harmonics, which are emitted around the maximum of the pulse. This means that the e-HHG harmonics can be identified through their spectral shift. Last, we note that the long lasting narrowband xFID emission from the excited states is visible between 9 and 14 eV at later times, in agreement with previous observations \cite{camp15}. 
 
The spectral shift of the e-HHG radiation can be understood within the Lewenstein model in which the instantaneous frequency of the \textit{q}th order harmonic, $\Omega_i ^ q$, can be expressed as $ \Omega_i ^ q = q \cdot \omega _i ^ {IR} + \alpha_q ^ {s,l} \cdot \partial I_0(t)/\partial t $, where $\omega _i ^ {IR}$ is the instantaneous frequency of the driving laser. If the harmonic is emitted during the falling edge of the pulse, $\partial I(r,t)/\partial t $ is negative and the long trajectory contribution to the harmonic line is redshifted. As the peak laser intensity increases, $\partial I(r,t)/\partial t$ is more negative and the redshift becomes more prominent, in good agreement with the experimental observations. Quantitatively, the energy shift is given by $-4\alpha I_0 c e^{(-2c^2)}/\tau$ where $c$ is the time at which recombination occurs in unit of pulse duration (i.e. at the time $c \tau$). SFA calculations provide $\alpha \approx 12 \times 10^{-14}$ cm$^2$/W for cutoff harmonics in the e-HHG process. Using the emission time of 4 fs for e-HHG (i.e. $c \approx 0.5$), we obtain a slope of the energy shift of $-17.5$ a.u. ($-1.35 \times 10^{-11}$ meV/cm$^2$) which is in good agreement with the experiment ($-1.2 \times 10^{-11}$ meV/cm$^2$) and in close agreement with TDSE which is known to overestimate $\alpha$ coefficients. 

\begin{figure}
\begin{center}
\includegraphics[width=8 cm,keepaspectratio=true]{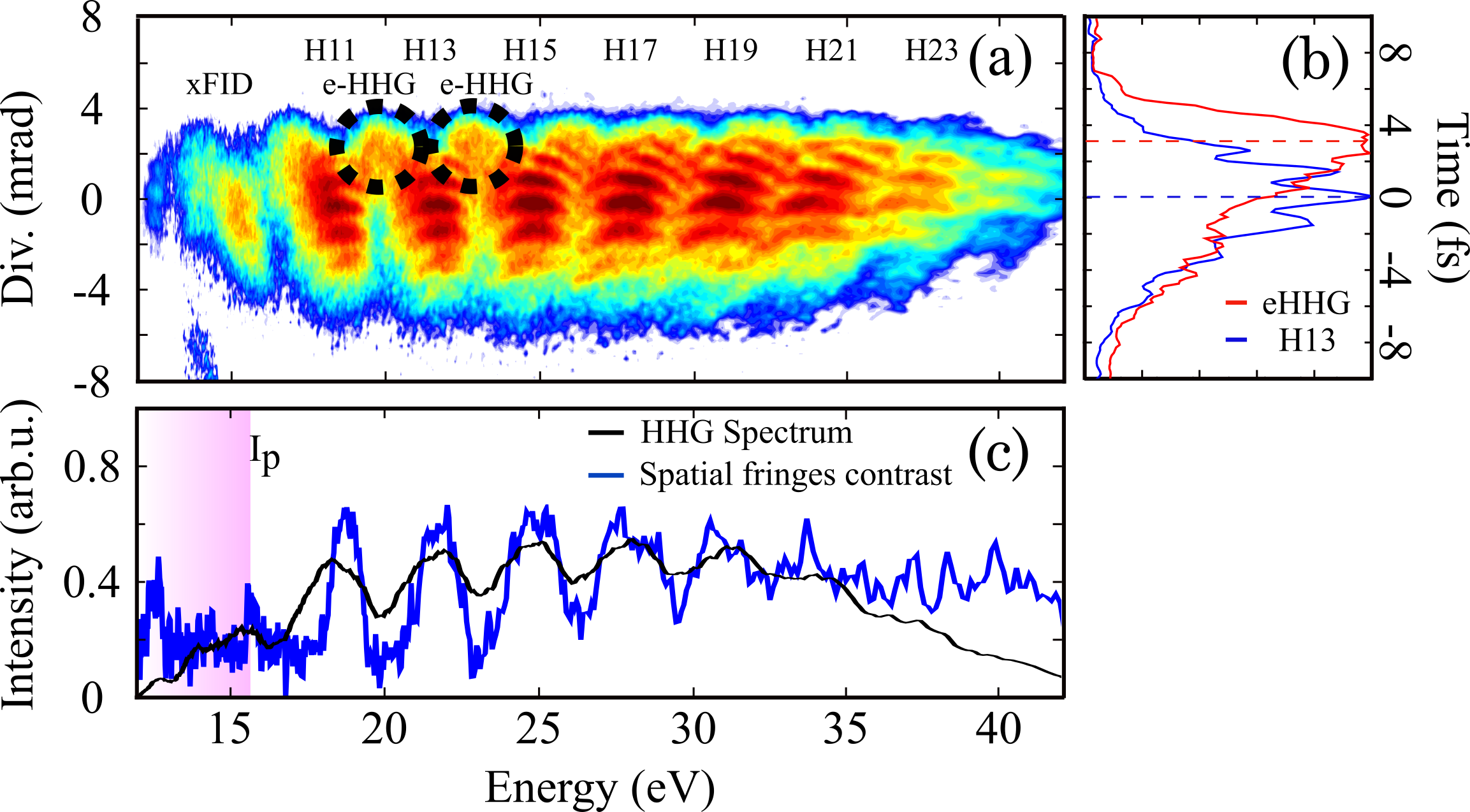}
\caption{Time-frequency mapping of HHG using the attosecond lighthouse technique. (a) Spatially-resolved harmonic spectrum driven by a spatially chirped laser pulse. (b) Spectrally integrated harmonic spatial profile for e-HHG (in red) and for H17 (in blue). The interfringe was used to transform the spatial axis into a time axis. (1 interfringe = 1.3 fs) (c) Spatially integrated harmonic spectrum (black) and contrast of the spatial fringes (blue). }
\label{fig4}
\end{center}
\end{figure}

The spatio-spectral measurements of the harmonic signal for different laser intensities have enabled us to identify a new component in the HHG process, associated to HHG from an excited state. This e-HHG mechanism is clearly distinguishable by time-frequency analysis of the TDSE dipole. In an attempt to experimentally observe this effect, we used the attosecond lighthouse technique, which consists in spatially separating different attosecond pulses from an attosecond pulse train through an ultrafast rotation of the laser wavefront at the focus \cite{vincenti12}. Each pulse of the attosecond pulse train is emitted in the direction orthogonal to the instantaneous wavefront. If the wavefront rotates fast enough, then two consecutive attosecond pulses are sent to different directions and separated on the detector in the far-field. The attosecond lighthouse thus maps the emission time of the harmonics onto their propagation direction. Up to now, this technique has only been used with laser sources whose carrier-
envelop phase (CEP) was stabilized, because changes of the CEP shift the position of the attosecond pulses on the detector \cite{quere14}. To avoid using a CEP-stabilized laser source, we performed single-shot acquisitions of the harmonic spectrum, with a 1 ms exposure time on the CCD camera. Clear fringes appear along the vertical dimension of the detector. From one laser shot to the next, the fringe pattern is found to shift, reflecting the fluctuations of the CEP. We recorded a series of images in these conditions, and tagged the CEP by measuring the phase of the spatial fringes by Fourier transform. We sorted the resulting images in CEP bins with -300 mrad $<$ $\phi_{CEP}$ $<$  300 mrad acceptance. 

Figure \ref{fig4}(a) shows the spatially streaked harmonic spectrum obtained using the attosecond lighthouse and averaged over 10 laser shots with the same CEP ($\pm$ 300 mrad). Well separated horizontal fringes are visible. They shift vertically as the CEP changes. Note that this is the first demonstration of single-shot CEP tagging of a CEP unstabilized laser using attosecond lighthouse. The level of sub-cycle temporal confinement of the harmonic emission can be evaluated by measuring the contrast of the spatial fringe pattern, using Fourier analysis. The results are shown in Fig. \ref{fig4}(b). Above threshold, the harmonics exhibit significant spatial fringe contrast, which is a signature of the sub-cycle (attosecond) confinement of their emission. On the other hand, this contrast falls down drastically below the ionization threshold, where xFID emission occur. As the photon energy further decreases, the contrast increases again around the non-resonant H9. The spatial fringe visible for below- and above-
threshold HHG is a signature of their attosecond pulse train temporal profile. The fact that spatial fringes are not observed for the xFID indicates that the emission is not confined on the attosecond time-scale.

A closer look at the lower energy part of the spectrum reveals that there are indeed two spatially separated components to the harmonic emission: a main comb, centered around the laser propagation axis, which shows well contrasted fringes, characteristic of the emission of well-confined attosecond pulses (see the cut of H13 in Fig \ref{fig4}(b)); a secondary, spectrally shifted comb, showing no spatial fringes and centered up on the detector. The spatial shift of the second component is the signature of a time delay in the emission, which can be quantified by calibrating the space-time mapping onto the detector using the fringe spacing as a time unit of 1.3 fs. The secondary component is found to be maximum around 4 fs, which is approximately the temporal delay between the conventional and e-HHG observed in the Gabor analysis of the TDSE. The attosecond lighthouse thus enables us to resolve experimentally the delay in the e-HHG emission. This measurement paves the way towards extending the scope of the 
attosecond lighthouse technique, from a way to generate isolated attosecond pulses to a metrology tool that enables the measurement of ultrafast electronic dynamics during HHG. 

In conclusion, we have demonstrated that excited states can emit XUV radiation through Free Induction Decay or by a new e-HHG process, where electrons are ionized from an excited state and recombine to the ground state. Using the attosecond lighthouse, we demonstrated that the xFID does not exhibit attosecond structure while the e-HHG is emitted only during the falling edge of the laser pulse. 
The e-HHG process is  likely to occur in many conditions, opening a broad range of applications: investigation of resonances in the continuum, broadband spectroscopic experiments using mid-infrared lasers, study of excited states in HHG from solids \cite{ghimire14,higuchi14,vampa15}. All these experiments could be time-resolved by using a pump pulse to photoexcite the sample and a probe pulse for e-HHG. Additional control parameters could be employed to increase the dimensionality of the measurements. For instance circularly polarized pump pulses would create ring currents, which could be mapped through the ellipticity dependence of the e-HHG, reflecting the momentum distribution of the ionized electrons \cite{Xie08}.

We thank V. Blanchet, B. Fabre, F. Qu\'er\'e, J. Mauritsson, B. Piraux and B. Pons for fruitful discussion, R. Bouillaud and L. Merzeau for technical assistance, and E. Constant and E. M\'evel for providing experimental apparatus. We acknowledge financial support of the European Union (LASERLAB-EUROPE II 228334 and LASERLAB-EUROPE 284464), the French National Research Agency (ANR), through ANR-MISFITS and IdEx Bordeaux – LAPHIA (ANR-10-IDEX-03-02, and the 
Vanier Scholarship. At LSU, this work was supported by the NSF under Grant No. PHY-1403236. Computer time was provided by the computing facilities MCIA.

\end{document}